# Analysis of Provincial Export Performance in Türkiye: A Spectral Clustering Approach

## Türkiye'de İllerin İhracat Performansının Analizi: Spektral Kümeleme Yaklaşımı

**Emre AKUSTA[1]**

[1]Dr. Öğr. Üyesi, Kırklareli Üniversitesi, İktisadi ve İdari Bilimler Fakültesi, İktisat Bölümü, Kırklareli, emre.akusta@klu.edu.tr,  orcid.org/0000-0002-6147-5443

*Araştırma Makalesi/Research Article*

**ABSTRACT**
This study analyzes and clusters Türkiye's 81 provinces based on their export performance. Import, export and net export data for 2023 are used in this study. In addition, exchange rate-adjusted versions of the data were also included to eliminate the effects of exchange rate fluctuations. Spectral clustering method is used to group the export performance of cities. The optimum number of clusters was determined by the Eigen-Gap method. The Silhouette coefficient method was used to evaluate the clustering performance. As a result of the analysis, it was determined that the data set was optimally separated into 3 clusters. Spectral-clustering analysis based on export performance showed that 42% of the provinces are in the "Low", 33% in the "Medium" and 25% in the "High" export performance category. In terms of import performance, 44%, 33%, 33%, and 22% of the provinces are in the "Medium", "High", and "Low" categories, respectively. In terms of net exports, 38, 35% and 27% of the provinces are in the "Low", "Medium" and "High" net export performance categories, respectively. İzmir has the highest net export performance, while İstanbul has the lowest.

**ÖZ**
Bu çalışma, Türkiye'nin 81 ilini ihracat performanslarına göre analiz etmekte ve kümelendirmektedir. Çalışmada 2023 yılı ithalat, ihracat ve net ihracat verileri kullanılmıştır. Ayrıca, döviz kuru dalgalanmalarının etkileri ortadan kaldırmak için verilerin döviz kurundan arındırılmış versiyonları da analize dahil edilmiştir. Çalışmada şehirlerin ihracat performanslarını gruplamak için spektral kümeleme yöntemi kullanılmıştır. Son olarak, optimum küme sayısı ise Eigen-Gap yöntemi ile belirlenmiştir. Kümeleme performansını değerlendirmek için siluet katsayısı yöntemi kullanılmıştır. Yapılan analiz sonucunda veri setinin optimal olarak 3 kümeye ayrıldığı tespit edilmiştir. İhracat performansına dayalı spektral-kümeleme analizi, illerin %42'sinin "Düşük", %33'ünün "Orta" ve %25'inin "Yüksek" ihracat performansı kategorisinde yer aldığını göstermiştir. İthalat performansı açısından illerin %44'ü "Orta", %33'ü "Yüksek" ve %22'si "Düşük" ihracat performansı kategorisinde yer almaktadır. Net ihracat açısından ise illerin %38'i "Düşük", %35'i "Orta" ve %27'si "Yüksek" net ihracat performansı kategorisinde yer almaktadır. Analiz sonuçlarına göre, İzmir en yüksek net ihracat performansına sahipken, İstanbul en düşük net ihracat performansına sahip ildir.

## 1. Introduction

Globalization is an important process that determines and transforms the dynamics of today's world economy. Technological developments, increased foreign capital inflows and growth in international trade have increased the interaction between countries. This process refers to the integration of world economies, free market economy, increased economic cooperation and integration in a single market. The historical development of globalization dates back to the 1870-1930 period when capitalism began to rise. With the collapse of the Bretton Woods system and the dissolution of the Soviet Union, the concept of globalization has become more widespread (Slobodian, 2015). The new world order established after the Second World War focused on the liberalization of international trade and sustainable development. International organizations such as the IMF, the World Bank and the World Trade Organization played important roles in this process. Globalization is a process that continues today and has significant effects on the world economy. International trade, international factor mobility and direct capital investments have developed and increased. After 1990, with the collapse of the Eastern Bloc, the countries included in this process entered a process of economic transformation. As a result, they have contributed to the expansion and acceleration of the globalization process (Kriesler and Nevile, 2016; Khan et al., 2023).

The economic dimension of globalization affects economic growth through international trade, financial integration, international labor mobility and technological changes. In this process, trade liberalization includes the removal of barriers to trade and the liberalization of international trade (Osada, 2015; Zahonogo, 2018). Financial liberalization, however, involves the harmonization of national financial markets with international markets by removing controls and restrictions on financial markets (Khan et al., 2023). In this way, it is aimed to ensure more efficient operation of institutions and specialization in production by reducing the cost of capital.

Economic globalization supports economic growth by liberalizing international trade and promoting financial integration. International trade was first proposed by Adam Smith and David Ricardo. In these theories, international trade was recognized as one of the important positive factors of economic growth. According to Smith and Ricardo, as countries produce specialized goods, foreign trade or trade openness is an important factor in increasing countries' incomes. New growth theories also recognize that open economies grow faster than closed economies. It also shows that trade openness increases economic growth by expanding the scale of spillovers (Romer, 1990). Within this theoretical framework, trade openness plays a vital role in the process of economic growth in the globalizing world economy, especially in developing countries (Hye and Lau, 2015).

The positive effects of globalization on economic growth include increased international trade, increased foreign capital investments, access to new technologies, improved economic output, increased export revenues through diversification of exported products, creation of new employment opportunities, and improved trade and budget balances (Dreher, 2006; Sharma et al., 2019). In addition, as production activities develop internationally in the globalization process, international labor migration increases. This reduces the costs of large companies. Moreover, globalization leads to improved access to technology, reduced costs and increased competition Li and Reuveny, (2003). The relationship between globalization and economic growth is a dynamic and multidimensional process. Therefore, globalization brings with it various challenges and negative effects as well as factors that support economic growth. The negative effects of globalization include increased inequality in the labor market, volatility of exchange rates, economic fragility, inability of some countries to adapt to increased competition, and reduced social spending. In order to benefit from the advantages of globalization, countries need to have strong economic structures and develop policies to adapt to international competition. Moreover, in order to better comprehend the effects of globalization on economic growth, it is crucial to quantitatively measure and analyze these effects (Ying et al. 2014; Gurgul and Lach, 2014). The complexity of the effects of globalization on economic growth increases the role of trade components such as exports and imports. International trade liberalization and financial integration directly affect the economic performance of countries. In this process, the relationships between exports, imports and economic growth can manifest in different directions. The positive effects of globalization on trade include higher export revenues, new employment opportunities and economic development. However, it is clear that these effects do not occur in the same way for all countries and periods. Therefore, the relationship between globalization, foreign trade and economic growth needs to be examined in more depth.

Globalization has become an important factor in the process of economic growth for developing countries such as Türkiye. This is because it has accelerated Türkiye's integration with the world economy and led to significant increases in export and import volumes. In particular, the reduction of customs barriers, liberalization of foreign capital flows and the promotion of international trade have further integrated Türkiye into global markets. It also realized growth opportunities in the industrial and service sectors. In this period, Türkiye's foreign trade volume grew rapidly, and export growth became the engine of economic growth. However, the effects of globalization on economic growth and development are different for each country and region. Türkiye's case is an example of these differences (Baş, 2009; Arslan, 2013).

Foreign trade, one of the elements of globalization that trigger economic growth in Türkiye, has also caused regional development imbalances. Large cities such as Istanbul, Izmir and Bursa have gained advantages in terms of economic growth and development (Oğul, 2021). These provinces have become the pioneers of the Turkish economy with their industrial production, export capacity and potential to attract foreign investment. However, this has also led to a deepening of interregional development gaps in Türkiye. For example, some provinces in the interior of Anatolia have not benefited sufficiently from globalization and foreign trade. The potential for economic growth in these regions has remained limited. This imbalance further deepens socioeconomic disparities and emphasizes the importance of development policies.

The differentiation of globalization across provinces is an obstacle to the sustainability of economic growth. Türkiye needs to ensure balanced regional development while integrating into global markets. Intensive commercial activities and industrial investments in major cities contribute to the growth of the national economy. However, rural provinces and low-capacity cities do not benefit sufficiently from this process. If globalization and trade do not create balanced development across the country, these imbalances can lead to economic and social problems in the long run. In this regard, provinces with low export performance need to be supported, integrated into global trade and local firms need to have access to international markets. Developing Türkiye's export potential and addressing trade imbalances between provinces requires a comprehensive strategic planning. In this planning, it is crucial to identify provinces with low export capacity, improve the export skills of firms in these provinces, increase infrastructure investments and facilitate access to international markets. Moreover, the production capacities of underdeveloped regions of Anatolia should be increased and these regions should be integrated into the global economy.

In conclusion, the effects of globalization and foreign trade on Türkiye's economic growth should be analyzed in more depth, considering regional differences. Türkiye's integration into world trade has not been achieved in all provinces. Therefore, identifying provinces with low export capacity will increase Türkiye's competitiveness in the global economy and ensure sustainable growth. Therefore, this study investigates the export performance of provinces in Türkiye and categorizes provinces based on their export performance. Therefore, the main problematic of this study is to identify provinces with low export capacity and develop policy recommendations to overcome this problem.

This study can contribute to the literature in at least 6 ways: (1) To the best of our knowledge, there is no study analyzing Türkiye's provinces based on their foreign trade performance. This study aims to fill this gap in the literature. (2) A real time and current perspective are provided by using the most recent data for the year 2023. (3) The data are adjusted to the dollar to eliminate the effects of exchange rate fluctuations. Thus, the effects of exchange rates on export performance are minimized. While other studies usually ignore this factor, this study comprehensively addresses exchange rate effects. (4) Methodological diversity is provided by using various statistical techniques such as spectral algorithm, Eigen-gap analysis and silhouette coefficient. (5) This classification using spectral clustering method identifies the export potential of different regions. This allows the development of policies that consider regional differences. (6) The visualization of the research findings facilitates the comprehension of the results and is an example of an innovative approach in the presentation of this type of analysis. Instead of the complex tables common in other studies, this study visualizes the findings in a more understandable format for both academics and policy makers. This also ensures an effective presentation of the results.

## 2. Literature Review

This study examines the export performance of provinces in Türkiye in the context of globalization and economic growth. Therefore, the literature review aims to examine the issues of globalization, foreign trade and economic growth. For this purpose, the literature review is divided into three main thematic sections. First, the theoretical literature is reviewed. In this section, the relationship between economic growth, foreign trade and globalization is examined. Second, studies focusing on the effects of globalization on economic growth are reviewed. This section discusses academic views on the effects of the globalization process through various mechanisms such as economic integration, capital flows, technology transfer and international labor mobility. Third, the literature on the effects of foreign trade - especially exports and imports - on economic growth is reviewed. This section reviews academic studies on the direct and indirect effects of export and import activities on economic performance, how they contribute at the national and international level, and the causal relationship between them and economic growth. Finally, studies that classify provinces in Türkiye by various indicators are presented. These studies analyze the regional status of economic, social and foreign trade activities at the local level.

The first part of the literature review examines the theoretical literature. The argument that exports are an important driver of economic growth is widely debated in the literature. Studies on this issue have not reached a common conclusion on the direction of the causal relationship between exports and economic growth (Bilgin and Şahbaz, 2009). The causal relationship between exports and economic growth in open economies can be explained in different approaches. First, the "export-led growth" hypothesis argues that exports positively affect economic growth. Under this hypothesis, increased exports allow shifting resources to more productive sectors, lead to productivity gains and encourage the adoption of new technologies through international competition (Van den Berg and Lewer, 2015; Grossman and Helpman, 1993). Moreover, increased exports enhance import capacity, making capital and intermediate goods more readily available. Second, the "growth-led exports" hypothesis posits that economic growth boosts exports. Under this hypothesis, economic growth facilitates the adoption of new technologies and leads to productivity gains, thereby increasing exports by gaining a competitive advantage in international markets (Giles and Williams, 2000). The third hypothesis argues that there is a two-way causal relationship between exports and economic growth. This hypothesis suggests that while higher exports stimulate economic growth, higher income levels may also increase trade, hence a two-way interaction. Finally, it is also believed that there may not be a causal relationship between exports and economic growth.

It should not be ignored that imports also play an important role in the relationship between exports and economic growth. The impact of foreign trade on economic growth is analyzed within the framework of export-led growth and import-led growth models. Based on the export-led growth model, domestic productivity increases and indirect effects of exports have positive effects on economic growth (Taştan, 2010; Yıldız and Berber, 2011; Akkaş and Öztürk, 2016; Pata, 2017). Based on the import-led growth model, imports contribute to economic growth by providing the necessary inputs for production (Yurdakul and Aydın, 2018). Endogenous growth theories suggest that imports are an important channel for new technology transfers and access to higher quality capital goods and intermediate goods. In this regard, imports can directly affect economic growth and this relationship can be explained by the "import-led growth" hypothesis (Coe and Helpman, 1995; Awokuse, 2008). Since it is not clear which of these models is theoretically valid, the relationship between trade and growth is an empirical problem. One of the motivations for research in this area is the lack of consensus.

As one of the determinants of economic growth, exports play a critical role in developing countries in the face of limited foreign exchange reserves and difficulties in obtaining resources from international financial markets (Şimşek and Kadılar, 2010). The effects of exports on economic growth are realized through many positive channels such as increasing competition, providing access to international markets, transferring know-how, encouraging large-scale entrepreneurship, increasing productivity and acquiring new technologies (Şimşek, 2003). Given the positive impact of exports on economic growth, it is vital for countries to maximize their export potential for economic growth.

The second part of the literature review examines the studies on globalization and economic growth. Among these studies, Chang and Lee (2010) found a unidirectional causality relationship from economic and social globalization to economic growth in their study on OECD countries. Quinn, Schindler and Toyoda (2011) conducted a dynamic panel data analysis on 189 developed and developing countries and concluded that economic globalization increases economic growth. Sakyi (2011), in his study on 31 Sub-Saharan African countries, found that economic globalization positively affects economic growth in the long run. Villaverde and Maza (2011) analyzed the data of 101 countries for the period 1970-2005 and showed that economic, social and political globalization has a positive impact on economic growth. Similarly, Ray (2012), in his study on India, found that there is a reciprocal causality between globalization and economic growth. Ali and Imai (2013) found a positive relationship between economic globalization and economic growth in their panel data analysis on 41 countries. Chang, Berdiev and Lee (2013), in their study on Azerbaijan, Georgia, Russia and Türkiye, found that there is a positive relationship between globalization and economic growth and that higher energy exports lead to higher growth rates. Doğan (2013), in his study on Türkiye, found a bidirectional causality relationship between economic globalization, social globalization and general globalization and economic growth.

Samimi and Jenatabedi (2014), in their study on the member countries of the Organization of Islamic Cooperation, found that the relationship between economic globalization and economic growth is positive in high-income countries and negative in low-income countries. Ying et al. (2014), in their study on the Association of Southeast Asian Countries, stated that economic globalization has a positive effect on economic growth and a negative effect on social and political globalization. Hayaloğlu, Kalaycı and Artan (2015), in their study on 91 countries, found that globalization increases economic growth, but social and economic globalization decreases economic growth. Kılıç (2015), in a study on 74 developing countries, found a bidirectional causality relationship from economic globalization to economic growth and a unidirectional causality relationship from political and social globalization to economic growth. Chu, Chang and Sagafi-Nejad (2016), on OECD countries and China, find a causality relationship from globalization to economic growth for the Netherlands and the United Kingdom, and from economic growth to globalization for the United States. Doğan and Can (2016), on South Korea, find a positive relationship between economic, social and political globalization and economic growth. Gözgör and Can (2017), in their study on 139 countries, concluded that economic globalization positively affects economic growth only in upper and middle-income economies. Kaurin and Simic (2017), in their study on Central and Eastern European countries, found that general globalization has a significant positive effect on growth, while social and political globalization is statistically insignificant.

Barış and Barış (2018), in their study on 28 European Union countries, found that globalization positively affects economic growth. Çelik and Ünsür (2020), in their study on 88 countries, found a bidirectional causality relationship between economic growth and economic, social and technological globalization, and a unidirectional causality relationship from economic growth to general and political globalization. Saygın (2021), in his study on E7 countries, found that economic and political globalization contributed positively to economic growth, while the effect of social globalization on economic growth was statistically insignificant. Çadırcı and Kaya (2022), in their study on Organization for Economic Cooperation and Development countries, found a unidirectional causality relationship between economic globalization and economic growth and a bidirectional causality relationship between social globalization and economic growth. Günay and Günsoy (2022), in their study on Sub-Saharan African countries, found that general economic and social globalization have a positive effect on economic growth, while social globalization has a negative effect.

The third part of the literature review examines studies on foreign trade and economic growth. The relationships between exports, imports and economic growth have been comprehensively analyzed with various methods over different periods and countries. Among these studies, Bilgin and Şahbaz (2009) and Dereli (2018) examined the long-run relationship between exports and economic growth in Türkiye. Bilgin and Şahbaz (2009) find a bidirectional causality relationship between exports and growth in the short run and a unidirectional causality relationship in the long run. Dereli (2018) emphasized that Türkiye has realized an import-led growth and imports of high-tech products play an important role in economic growth. Gül and Kamacı (2012) tested the impact of foreign trade on growth in developed and developing countries with panel data analysis and found a causality relationship from imports and exports to growth. Kar, Nazlıoğlu and Ağır (2014), in their study on Türkiye, find a bidirectional causality relationship between economic growth and trade openness. Korkmaz and Aydın (2015) analyzed the effects of exports and imports on economic growth in Türkiye. The results show that there is a bidirectional causality relationship between imports and economic growth.

Uçan and Koçak (2014) and Sultanuzzaman, Fan, Mohamued, Hossain and Islam (2019) investigated the long-run effects of exports and imports on economic growth in various countries. Uçan and Koçak (2014) find a long-run relationship between growth and exports and imports. Sultanuzzaman et al. (2019) emphasized the positive impact of exports and technology on economic growth in Asian economies. Moreover, Assaf and Abdulrazag (2015) and Aytaç (2017) investigated the export-led growth hypothesis. Assaf and Abdulrazag (2015) found that exports support growth in Jordan in the short and long run. Aytaç (2017) finds that there is a unidirectional causality relationship between economic growth and exports in Türkiye and growth positively affects exports. Guntukula (2018) and İzgi and Yılmaz (2018) examined the long-run relationship between exports, imports and economic growth in India and Türkiye. Guntukula (2018) finds a causal relationship between exports and imports and economic growth in India and emphasizes the importance of export promotion strategies for sustainable growth. İzgi and Yılmaz (2018) find a unidirectional causality relationship from exports to economic growth in Türkiye and emphasize the importance of policies to increase exports for growth. Tekbaş (2019), in his study on BRICS and Türkiye, found a bidirectional causality relationship between economic growth and trade openness.

Mosikarı and Eita (2020) found that an increase in exports supports growth in Namibia, while a decrease in exports damages economic growth. Similarly, Demirel and İşcan (2021) find long-run relationships between exports and economic growth in Türkiye and South Korea and find that exports positively affect economic growth. Kong, Peng, Ni, Jiang and Wang (2021) and Omoke and Charles (2021) investigated the relationship between foreign trade and economic growth in China and Nigeria, respectively. Kong et al. (2021) stated that foreign trade in China has a positive impact on economic growth in the long and short run and encourages capital formation. Omoke and Charles (2021), on the other hand, found that exports have a positive effect on growth in Nigeria, while imports have a negative effect. Finally, Morley (2022) emphasized that the balance between exports and imports is important for sustainable growth and found that exports support economic growth. In conclusion, studies in the literature show that exports generally have a positive effect on economic growth. However, the effects of imports on economic growth vary across countries and periods. Especially exports of high value-added products play an important role in economic growth. This is an important factor that should be taken into account in determining the economic policies and trade strategies of countries.

The last part of the literature review examines studies that cluster provinces in Türkiye in terms of various indicators. Various studies have been conducted on the classification and clustering of provinces in Türkiye. Especially understanding and evaluating the socioeconomic structure of provinces has an important position in the academic literature. Researchers have classified and compared the development levels of provinces and regions using various statistical methods and socioeconomic indicators. For example, studies such as Kılıç, Saraçlı and Kolukısaoğlu (2011), Çakmak and Örkçü (2016), Arı and Hüyüktepe (2019) and Özkubat and Selim (2019) analyzed the socioeconomic structure of provinces in Türkiye. Among these studies, Kılıç et al. (2011) revealed the similarities and differences of provinces through socioeconomic indicators, while Çakmak and Örkçü (2016) determined the efficiency levels of provinces in certain areas through data envelopment analysis. While Arı and Hüyüktepe (2019) examine the development levels of provinces from a broad perspective with various analysis methods, Özkubat and Selim (2019) emphasize the differences between regions by creating development indices based on certain indicators. Again, Servi and Erişoğlu (2020) group provinces by their development levels based on certain socioeconomic indicators, while Karadaş and Erilli (2023) prepare a general socioeconomic development map with gray cluster analysis.

In addition to these studies, provinces in Türkiye have also been classified in terms of various indicators. For example, there are studies clustering provinces based on health, livestock, quality of life, entrepreneurship and transportation indicators. Studies such as Çelik (2013) and Tekin (2015) classify Türkiye's provinces on the basis of health indicators, while Çelik (2015) analyzes the impact of these sectors on the regional economy using agriculture and livestock data. Kandemir (2018) clustered provinces according to their tourism potential. These studies reveal how cultural and touristic characteristics can be evaluated in regional analysis. Yılmaz and Uzgören (2014) and Sarıgül (2014) analyzed the effects of the banking sector on provinces with different methods. While Uzgören (2014) categorized provinces into six groups in terms of banking activities, Sarıgül (2014) categorized provinces into homogenous groups in terms of access to and utilization of banking services. Finally, Işık and Polatgil (2020) group provinces based on entrepreneurship activities, while İncekırık and Altın (2021) divide provinces into five clusters based on transportation indicators.

The findings of the literature review reveal the complexity of globalization, economic growth and foreign trade relations and the effects of these dynamics on the economic performance of countries. Moreover, the necessity of accurately assessing and investigating the export potential of countries is emphasized. Thus, in order for Türkiye to realize its foreign trade performance more effectively, the importance of investigating the export potential on a provincial basis has emerged. This approach offers the opportunity to utilize the country's overall export potential more effectively by taking into account regional differences and unique economic strengths. Therefore, this study examines Türkiye's provincial export performance in the context of globalization and economic growth.

## 3. Data and Methodology

### 3.1. Model Specification and Data

The main objective of this study is to investigate the export performance of provinces in Türkiye and to cluster them accordingly. The data set used in the study consists of import, export and net export data provided by the Turkish Statistical Institute (TurkStat). These data belong to the year 2023. In addition, exchange rate-adjusted versions of the data are also included to eliminate the effects of exchange rate fluctuations. The exchange rate adjustment is done to more accurately reflect the actual impact of import costs on exports. Spectral Clustering method is used to group the export performance of provinces. The optimum number of clusters is determined by the Eigen-Gap method. The Silhouette Coefficient method is used to evaluate the clustering performance. Finally, six maps showing the import, export and net export performance of each city were prepared to make the research findings more understandable. These maps include both raw and exchange rate-adjusted data and provide a visual representation of the economic performance of the cities. Descriptive statistics of the data set used in the study are shown in Table 1.

**Table 1.** Descriptive Statistics

| Variables | Description | Mean | Mdn. | S. D. | Min. | Max. | Source |
|---|---|---|---|---|---|---|---|
| Export | Thousand US $ | 3153558 | 304126 | 14251811 | 139 | 127097589 | TurkStat |
| Imports | Thousand US $ | 4008624 | 152298 | 152298 | 213 | 203362138 | TurkStat |
| Net exports | Thousand US $ | -855065 | 30170 | 30170 | -76264549 | 4245678 | TurkStat |
| Export-adjustment | Thousand US $ | 136213 | 13183 | 13183 | 5 | 5503134 | TurkStat |
| Import-adjustment | Thousand US $ | 175134 | 6701 | 6701 | 10 | 8882201 | TurkStat |
| Net exports-adjustment | Thousand US $ | -38920 | 1323 | 1323 | -3379066 | 183522 | TurkStat |

Note: S.D., Min, Max, Mdn, TurkStat, and adjusted denote standard deviation, minimum, maximum, median, Turkish Statistical Institute, and exchange rate adjustment respectively.

Table 1 presents descriptive statistics on the export, import and net export performance of provinces in Türkiye. Exports and imports are analyzed with raw and exchange rate-adjusted data. In both cases, the data have a highly right-skewed distribution, meaning that the means are well above the median values. This implies that some provinces with particularly high export or import values pull these averages up.

Export data shows that the average export value in the raw data is 3,153,558.2 thousand dollars, while the median value is 304,126.2 thousand dollars. This difference reveals that export values vary greatly across provinces, with some provinces realizing much higher exports than others. This is also true for the exchange rate-adjusted export data, where the average export value is 136,213.4 thousand dollars, but the median value is much lower at 13,183.4 thousand dollars. On the import side, the raw data show a mean value of $4,008,624.0 thousand and a median value of $152,298.1 thousand. There is a similar skewness and large variation in import values. In exchange rate-adjusted data, the average import value is $175,134.3 thousand dollars and the median is $6,701.9 thousand dollars. Net export values show how much more or less exports are than imports. In raw data, average net exports are negative at -855,065.8 thousand dollars, while in exchange rate adjusted data this value drops to -38,920.8 thousand dollars. In both cases there is a large difference between the minimum and maximum values, suggesting that some provinces differ significantly from others in terms of net exports.

### 3.2. Cluster Analysis: Spectral Clustering

Spectral clustering is a powerful method used to partition datasets into their natural groups. The Eigen-Gap methodology, which is a fundamental part of this method, is important for determining the optimal number of clusters. The first step in the process is to create a similarity matrix that expresses the similarities between data points. This matrix is usually calculated using Gaussian kernel or k-Nearest Neighbor techniques and presents the relationships between data points in a numerical format (Xia, Cao, Zhang and Li, 2009). The spectral clustering method is basically implemented in 6 steps (Wen and Li, 2019; Wen and Li, 2020).

*1. Data Preparation:* In this study, the data set is defined as $X = \{x_1, x_2, ..., x_n\}$. $x_i$ is a $d$-dimensional data point. In the pre-processing stage of the dataset, missing values were imputed and the data was normalized.

*2. Similarity Matrix Creation:* A similarity matrix $S$ was created to represent the similarities of the data points. Gaussian kernel function was used to create this matrix:

$$S_{ij} = exp\left(\frac{\|x_i - x_j\|^2}{2_{\sigma^2}}\right) \tag{1}$$

The parameter σ in Equation 1 is used to scale the distances between data points. The similarity matrix forms a symmetric matrix representing the relationships between the data.

*3. Laplace Matrix and Eigen Decomposition:* Based on the similarity matrix, the Laplacian matrix is calculated in the next step of spectral clustering. The Laplacian matrix can take two basic forms: standard and normalized. In the standard form, the Laplacian is calculated as $L = D - S$, while in the normalized form $L_{sym} = D^{-1/2}SD^{-1/2}$ is used. This is a diagonal matrix consisting of the row sums of the similarity matrix. Next, the eigenvalues and eigenvectors of the Laplacian matrix are calculated (Huang, Chuang and Chen, 2012).

*4. Clustering:* The first $k$ eigen vector (ordered from smallest to largest) obtained as a result of eigen decomposition is put together to form a matrix $U$. The rows of this matrix represent the representations of each data point in the new $k$-dimensional space. A classical $k$-Means algorithm is applied on these representations to cluster the data points (Liu, Lin, Yan, Sun, Yu and Ma, 2012; Ji, Salzmann and Li, 2014).

$$y_i = \arg min_j \|U_i - \mu_j\|^2 \tag{2}$$

In Equation 2, $y_i$, represents the cluster label to which the $i$-th data point belongs and $\mu_j$ represents the centers of the $k$-Means algorithm. As a result of this step, each data point in the dataset is assigned to a specific cluster. These clusters form meaningful groups based on the similarities of the data points and reveal the intrinsic properties of the data structure.

*5. Determination of the Number of Clusters:* The Eigen-Gap method was used to determine the number of clusters. This method is an important part of the spectral clustering process and is an effective approach for selecting the optimal number of clusters. The Eigen-Gap method is based on the principle of ranking the eigen values according to their magnitudes and examining the differences between these values. First, the eigenvalues of the normalized Laplace matrix are calculated and arranged in ascending order: $\lambda_1 \leq \lambda_2 \leq \cdots \leq \lambda_n$. Afterwards, the differences between consecutive eigenvalues are calculated: $\Delta_i = \lambda_{i+1} - \lambda_i$. These differences help to identify the natural cluster structure in the dataset. The location of the largest eigen difference marks the point where the data points are clearly separated. The number of eigenvalues before this difference indicates the optimal number of clusters. Therefore, the value $k$ with the largest difference is chosen as the optimal number of clusters: $k = \arg max_i \Delta_i$ (White and Smyth, 2005; Fan, Tu, Zhang, Zhao and Zhang, 2022).

*6. Evaluation Metrics:* Various metrics such as Silhouette Coefficient, Rand Index and V-Measure are used to evaluate clustering performance. Silhouette Coefficient is used in this study. Because Silhouette coefficient is used to evaluate the effectiveness of a clustering solution and measures the cohesion of each data point within its cluster and its separation from other clusters. The Silhouette score takes values between -1 and +1. If a data point's score is close to +1, it is very well placed within its cluster and well separated from the nearest cluster. If the score is close to -1, the point is misplaced in its cluster. When the score is close to 0, the data point is on the boundary between clusters. The Silhouette coefficient is calculated by the formula in Equation 3 (Rousseeuw, 1987):

$$s(i) = \frac{b(i) - a(i)}{\max[a(i), b(i)]} \tag{3}$$

The function $a(i)$ in Equation 3 is the average distance of data point $i$. to other points in its cluster. The $b(i)$ function is the average distance of data point $i$. from the next closest cluster. That is, $a(i)$ represents intra-cluster agreement and $b(i)$ represents inter-cluster separation.

In the spectral clustering method, unlike the $k$-Means algorithm, the "elbow" method is not used directly. The "elbow" method used in the $k$-Means clustering algorithm is a frequently used technique for selecting the number of clusters for optimization (Maadooliat, Sun and Chen, 2018). This method is based on plotting the ratio of the within-cluster variance to the total variance on a graph and observing an "elbow" at a point where increasing the number of clusters does not reduce the total variance. In the $k$-Means method, this point is considered to be the point at which increasing the number of clusters no longer provides a significant reduction (Little, Maggioni and Murphy, 2020).

In spectral clustering, a different strategy is used to determine the number of clusters. Therefore, in this study, the Eigen-Gap method is used to determine the optimal number of clusters and the Silhouette coefficient method is used to evaluate the clustering performance. These two methods are among the most widely used and proven methodologies in spectral clustering analysis. The Eigen-Gap method detects large gaps between eigenvalues and shows how the dataset should be naturally separated. This method proposes a set of clusters that are compatible with the underlying geometric and topological properties of the data structure (Alzate and Suykens; 2008). The Silhouette coefficient evaluates each cluster separately and measures the separation between clusters and their internal consistency. The number of clusters with the highest Silhouette coefficient is considered to be the most representative of the data structure and therefore the preferred number of clusters. This method is used to directly compare the performance of the clustering algorithm with different cluster numbers and determine the optimal cluster configuration (Dinh, Fujinami and Huynh, 2019). A combination of both methods is used to obtain more robust and reliable clustering results.

## 4. Results and Discussion

### 4.1. Eigen-Gap Method Results

In this study, the natural cluster structure of the data was analyzed using spectral clustering method. The optimum number of clusters was determined by analyzing eigenvalues and Eigen-Gaps.

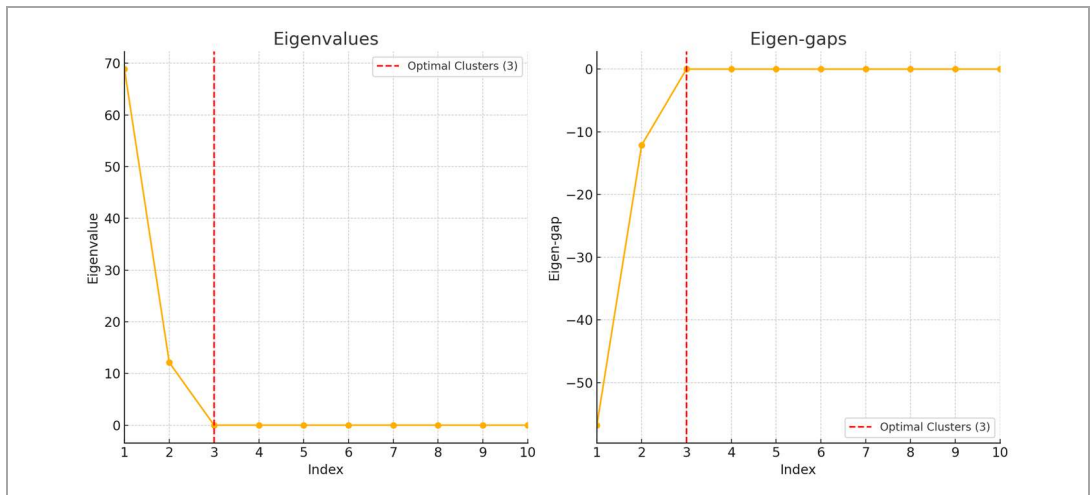

**Figure 1.** Eigen-Gap Method Results

Figure 1 shows the results of the Eigen-Gap method. In the eigenvalues plot, the y-axis shows the eigenvalues, and the x-axis shows the order of these eigenvalues. Higher eigenvalues indicate more densely connected clusters. In the eigenvalues plot, the first few eigenvalues are high and then decrease. Similarly, in the Eigenvalue ranges plot, the y-axis shows the difference between consecutive eigenvalues (eigenvalue range) and the x-axis shows the order of these differences. Eigenvalue ranges are important for identifying natural clusters in data. A large eigenvalue gap indicates that there is a clear separation in the data and that a natural cluster boundary exists at that point. In the Eigen-Gaps plot, the third index has a large eigenvalue gap. In both graphs, the red dashed line indicates the optimal number of clusters. This indicates that the data is divided into 3 natural clusters.

## 4.2. Silhouette Coefficient Method Results

In this study, the Silhouette coefficient method is used to evaluate clustering performance. The Silhouette coefficient is a metric that measures the difference between the similarity of each data point within its cluster and the closest other cluster. A high Silhouette coefficient indicates that data points cluster well within their cluster and are clearly separated from other clusters.

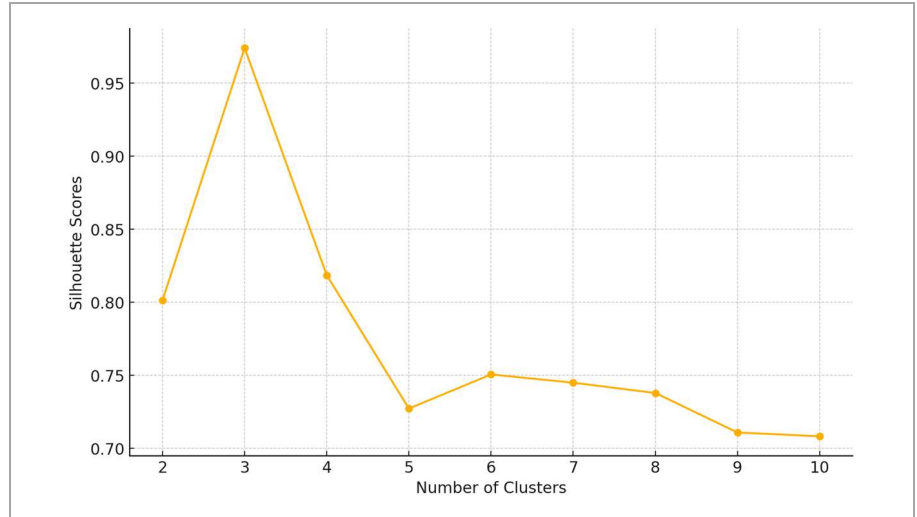

**Figure 2.** Silhouette Coefficient Method Results

Figure 2 shows the results of the Silhouette coefficient method. The y-axis shows the Silhouette scores and the x-axis shows the number of clusters. This graph helps to determine the optimal number of clusters by comparing the Silhouette scores for different cluster numbers. Silhouette coefficients range from 0 to 1, with higher values indicating better clustering quality. The results of the analysis show that the Silhouette coefficient in the two-cluster case is 0.80. In the three-cluster case, the Silhouette coefficient increased to 0.97. This high score indicates that the three-cluster structure is the most appropriate clustering for the data. At four clusters and beyond, the Silhouette coefficient started to decrease. The Silhouette coefficient decreased to 0.82 in cluster four and 0.72 in cluster five. In general, the Silhouette coefficients tend to decrease as the number of clusters increases. This decrease indicates that clustering quality decreases as the number of clusters increases. As a result, using both Eigen-Gap and Silhouette coefficient methods, it was determined that the data were optimally divided into 3 clusters.

## 4.3. Spectral-Clustering Analysis Results

In this study, we analyzed the eigenvalues of the Laplacian matrix with the Eigen-Gap method. By identifying the first large gap, we determined the number of clusters suggested by this gap. This step served as a basic guideline for identifying natural groups that reflect the internal structure of the dataset. We then performed a series of spectral clustering operations with this suggested number of clusters and calculated the Silhouette coefficients for each clustering iteration, which showed how well clusters were defined and how coherent items were within their clusters. As a result of these analyses, we determined the optimum number of clusters as 3. Therefore, the import, export and net export performances of the provinces were classified into 3 clusters as low, medium and high. The results of the analysis are shown in Figure 3.

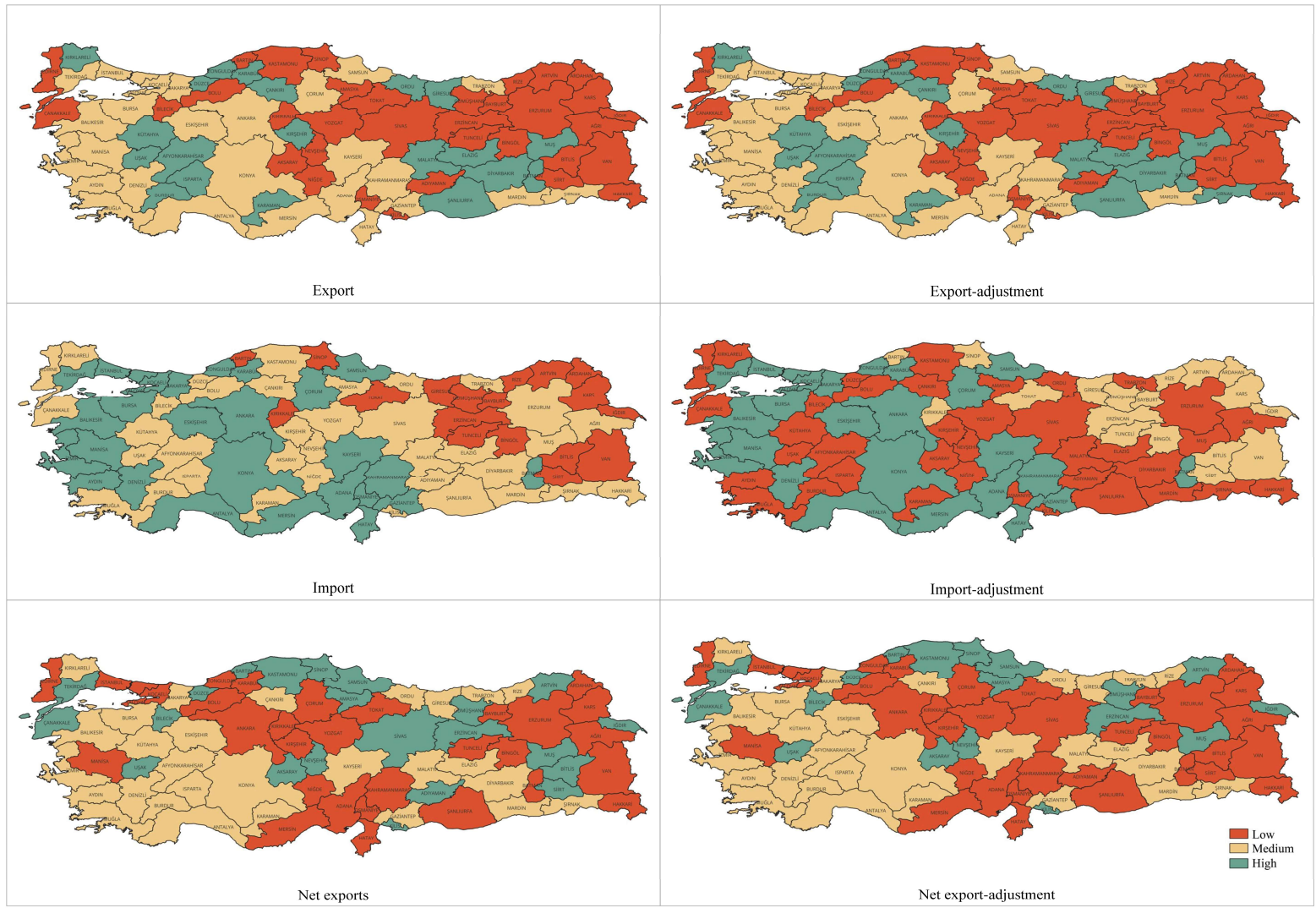

**Figure 3.** Spectral Clustering Analysis Results

133

Figure 3 shows the results of the Spectral-Clustering analysis with constant dollar and exchange rate adjusted data. The constant dollar data shows that 42% of provinces are in the "Low", 33% in the "Medium" and 25% in the "High" export performance category. This distribution indicates that Türkiye's export potential needs to be improved. Istanbul is the province with the highest exports with 127,097,589 thousand dollars. Bayburt stands out as the province with the lowest export performance with only 140 thousand dollars. Provinces with high export performance are generally economically developed, industrialized and highly populated cities. These provinces also constitute Türkiye's main trade and logistics centers. Provinces with medium and low export performance have economies based mostly on agriculture and small-scale industrial activities. These provinces need more economic support and investment.

In terms of import performance, 22% of the provinces in Türkiye are in the "Low", 44% in the "Medium" and 33% in the "High" import performance category. This distribution shows that Türkiye's import capacity is fairly balanced, but the high-importing provinces generally have larger economies. For example, Istanbul has the highest imports with 203,362,139 thousand dollars while Tunceli has the lowest imports with only 213 thousand dollars. The provinces with high import performance are economically developed regions, rich in population density and industrial activities. These provinces are also characterized as important trade centers. Provinces with medium and low import performance generally have smaller markets, less industrialized economies, and more local production. Net exports represent the difference between exports and imports and are an important indicator of trade balance. The results show that 38% of provinces are in the "Low", 35% in the "Medium" and 27% in the "High" net export performance category. This is a reflection of Türkiye's trade balance and regional economic disparities. İzmir stands out as the province with the highest net export performance with 4,245,679 thousand dollars. On the other hand, Istanbul has the lowest net export value with -76,264,549 thousand dollars. İzmir's high net exports are due to the fact that the city's exports exceed its imports, indicating that it is a region that provides foreign currency inflow economically. On the other hand, Istanbul's low net exports are due to the city's high import needs, reflecting its large industrial capacity and high consumption requirements.

Clustering results using exchange rate-adjusted data show that 42%, 32% and 26% of the provinces in Türkiye are in the "Low", "Medium" and "High" export performance categories, respectively. This distribution shows that the export potential of the vast majority of provinces needs to be improved. Istanbul stands out as the province with the highest exchange rate adjusted exports at 5,503,134 thousand dollars. On the other hand, Bayburt has the lowest exports with only 5 thousand dollars. The high export value of provinces such as Istanbul, Izmir, Kocaeli, Bursa and Ankara is due to their strong connections in international markets and large industrial capacity. In contrast, low exporting provinces such as Bayburt, Tunceli, Kars, Ardahan and Bingöl are less industrialized regions with limited economic activity.

An analysis of the exchange rate-adjusted import performance of provinces in Türkiye reveals that 47% of provinces are in the "Low", 31% in the "High" and 22% in the "Medium" import performance category. This distribution shows that the majority of provinces have low import capacity and limited import dependency. Istanbul stands out as the province with the highest exchange rate adjusted imports with 8,882,201 thousand dollars. However, Ardahan is recorded as the province with the lowest imports with only 10 thousand dollars. The high import value of provinces such as Istanbul, Kocaeli, Ankara, Izmir and Bursa stems from the need to import a large number of products due to the city's large industrial capacity and high consumption requirements. Low importing provinces such as Ardahan, Tunceli, Bayburt, Gümüşhane and Kars are less populated and less industrialized regions.

An analysis of the exchange rate-adjusted net export performance of provinces in Türkiye demonstrates that 43% of provinces are in the "Low", 35% in the "Medium" and 22% in the "High" net export performance category. This distribution is a reflection of Türkiye's foreign trade balance and regional economic differences. İzmir stands out as the province with the highest exchange rate adjusted net exports with 183,522 thousand dollars. Meanwhile, Istanbul has the lowest net export value with -3,379,067 thousand dollars. The high net exports of provinces such as İzmir, Gaziantep, Sakarya, Bursa, Kayseri and Denizli are due to the fact that the province's exports are higher than its imports, indicating that it is a region that provides foreign currency inflow economically. On the other hand, the low net exports of provinces such as Istanbul, Kocaeli, Ankara, Hatay, Çorum and Adana stem from their high import needs, reflecting their large industrial capacities and high consumption requirements. Provinces with high exchange rate-adjusted net export performance are regions whose exports exceed their imports, thus generating more foreign exchange inflows economically. Provinces with medium and low net export performance, in turn, are provinces that need more imports and experience difficulties in terms of foreign trade balances. This situation is generally observed in provinces with a large industrial structure.

Constant dollar export values indicate that some of Türkiye's major cities, particularly Istanbul and Izmir, have high export capacities. An analysis of exchange rate-adjusted export values reveals that the export performance of these provinces is affected by exchange rate fluctuations. For instance, while Istanbul's exports are high in constant dollar terms, this performance declines when adjusted for exchange rates. This shows that provinces with high exports are more sensitive to changes in global markets and exchange rate movements. Moreover, especially in large industrial and consumption centers such as Istanbul, import values in constant dollars are quite high. An analysis of exchange rate-adjusted import values reveals that the imports of these provinces are higher than those of other cities. This shows that the costs of imported products are directly affected by exchange rate changes and indicates that import dependency may lead to economic vulnerabilities. Finally, some provinces such as İzmir, Gaziantep, Sakarya, Bursa, Kayseri, Denizli and Konya exhibit positive net exports in constant dollar and exchange rate adjusted terms. However, provinces such as Istanbul, Kocaeli, Ankara, Hatay, Çorum and Adana are net importers even in constant dollar terms, and this is even more pronounced on an exchange rate-adjusted basis. This suggests that high import requirements do not sufficiently support local production capacity and that exchange rate fluctuations put pressure on net exports. Constant dollar and exchange rate-adjusted data reveal the effects of exchange rate fluctuations on the economic performance of provinces from different aspects. It is clear that exchange rate effects play a crucial role in economic performance, especially in provinces with high import and export volumes. Therefore, incorporating exchange rate fluctuations into economic planning and policy decisions can lead to more robust economic strategies. Such an approach can help Türkiye maintain its trade balance and promote sustainable economic growth.

## 5. Conclusions

There is a dynamic relationship between globalization, economic growth and foreign trade. If countries can evaluate their export potential well enough, they can realize economic growth with the opportunities provided by globalization. Therefore, countries, especially Türkiye, need to evaluate their export potential correctly. For this purpose, this study analyzes and clusters Türkiye's provinces based on their export performance. Import, export and net export data for 2023 are used in the study. In addition, exchange rate-adjusted versions of the data are also included to eliminate the effects of exchange rate fluctuations. Spectral clustering method is used to group the export performance of cities. The optimum number of clusters is determined by the Eigen-Gap method. The Silhouette coefficient method is used to evaluate the clustering performance.

The spectral-clustering analysis based on export performance shows that 42% of the provinces are in the "Low", 33% in the "Medium" and 25% in the "High" export performance category. Istanbul has the highest export performance in constant dollars, while Bayburt has the lowest export performance. In terms of import performance, 44% of provinces were in the "Medium", 33% in the "High" and 22% in the "Low" category. While Istanbul realized the highest imports in constant dollars, Tunceli was recorded as the province with the lowest import performance. In terms of net exports, 38% of provinces were in the "Low", 35% in the "Medium" and 27% in the "High" net export performance category. Izmir has the highest net exports, while Istanbul has the lowest net exports with a negative value. The differences between export, import and net export performances are even more pronounced when adjusted for the dollar exchange rate. While Istanbul has the highest exchange rate-adjusted exports and imports, it has the lowest net export performance. This reveals Istanbul's high need for imports and its economic structure that is sensitive to exchange rate fluctuations. İzmir, on the other hand, has a positive outlook in terms of both exports and net exports.

Exchange rate effects have clearly played a decisive role on economic performance, especially in provinces with high import and export volumes. This suggests that high import requirements do not adequately support local production capacity and exchange rate fluctuations put pressure on net exports. A number of policy recommendations have been developed for provinces that have not fully realized their export potential and to mitigate the impact of exchange rate fluctuations on foreign trade: (1) Investment in transportation, logistics and communication infrastructure should be increased in provinces with high export potential but inadequate infrastructure. Export processes can be accelerated by modernizing ports, expanding railway networks and increasing air cargo capacities. This will enable faster and more efficient transportation of products to international markets. Strengthening digital infrastructure will boost e-commerce and digital export capacity. (2) Invest in comprehensive technical education and vocational training programs to increase the skilled workforce. In particular, trainings on digital skills, international trade, logistics management and foreign trade law should be organized to bring the local workforce up to international standards. In addition, organize continuing education programs to inform local businesses on international trade and finance. (3) Provide credit and tax incentives to small businesses to increase their export capacity. These enterprises should be encouraged to enter export markets. They should be provided with digital marketing tools, support to participate in international fairs and consultancy services to help them establish a presence on online platforms. This support can take various forms, such as financing assistance, tax breaks and marketing support. (4) Exporting enterprises should be protected against exchange rate fluctuations. For this purpose, hedging and the use of derivatives should be encouraged. Local banks and financial institutions should provide exporters with appropriate hedging products and training on the effective use of these instruments. The development of new financial instruments to reduce exchange rate volatility could also be considered. (5) The government should review existing foreign trade agreements and create fairer and more balanced terms of trade. Existing agreements should be expanded and new ones signed to provide access to new markets. Strategic partnerships should be developed, especially in growing markets such as Africa, Southeast Asia and Latin America. This will provide exports with greater protection against exchange rate fluctuations. (6) Marketing campaigns should be organized to increase the recognition of local products in international markets. Special attention could be given to the promotion of products based on cultural heritage and products with geographical indications. Programs should be developed to encourage local businesses to participate in international fairs and virtual export fairs should be organized. In this way, the products of these provinces will be accepted in a wider market and export volumes will increase. (7) Encouraging the use of technology and innovation in export sectors increases productivity and product quality by modernizing production processes. R&D activities should be increased and international cooperation should be established. Promoting domestic technology increases the competitiveness of exporters in global markets and enables the export of high value-added products. Environmental sustainability should also be supported by investing in green energy technologies. (8) Encourage the production of high value-added products. In particular, incentive programs for strategic sectors such as technology, health, biotechnology and renewable energy should be expanded. High value-added products generally offer higher profit margins and are less price sensitive in international markets. This makes them less vulnerable to exchange rate fluctuations and increases export potential. (9) Develop regional export strategies that consider the economic structures and needs of different regions of Türkiye. In regions with intensive agriculture, investments can be made in the processing of agricultural products. Industrial centers, on contrary, should implement policies to increase advanced technology and production capacity. In this way, economic imbalances between provinces can be eliminated and export performance can be improved. (10) Sustainable production and green trade should be encouraged in the face of increasing environmental sensitivities in international markets. Exporters should be provided with incentives to be environmentally friendly and companies with low carbon footprints should be rewarded. In this way, Türkiye can gain a competitive advantage in exports by aligning with global sustainability goals.

This study has important implications. However, it also has some limitations. These limitations can be addressed in future studies. First, this study mainly uses import and export data. In future studies, an index can be developed with indicators affecting foreign trade and clustering can be done with this index. Second, Türkiye and its provinces were used as the sample in this study. In future studies, developed and developing countries can be determined as the sample. In this way, the results obtained can be compared. Third, spectral clustering, Eigen-Gap and Silhouette coefficient methods were used in this study. The results obtained in this study can be compared using different and/or hybrid techniques.

**Etik, Beyan ve Açıklamalar**

**1.** Etik Kurul izni ile ilgili;
☑**.** Bu çalışmanın yazarı, Etik Kurul İznine gerek olmadığını beyan etmektedir.
**2.** Bu çalışmanın yazarı, araştırma ve yayın etiği ilkelerine uyduklarını kabul etmektedir.
**3.** Bu çalışmanın yazarı kullanmış oldukları resim, şekil, fotoğraf ve benzeri belgelerin kullanımında tüm sorumlulukları kabul etmektedir.
**4.** Bu çalışmanın benzerlik raporu bulunmaktadır.